# Disordered transmission-line networks with and without parity symmetry

Tianshu Jiang and C. T. Chan*

*Department of Physics, Hong Kong University of Science and Technology, Clear Water Bay, Hong Kong, China*

* Correspondence address: phchan@ust.hk

## Abstract

Topological states are useful because they are robust against disorder and imperfection. In this study, we consider the effect of disorder and the breaking of parity symmetry on a topological network system in which the edge states are protected by Chern numbers. In the absence of periodicity, the local Chern number is adopted to characterize the topological features of the network. Our numerical results show that the local Chern number and the edge states are very robust against onsite disorder as long as the gap of the bulk state continuum remains open and survives even when the bulk band gap is closed. Breaking the parity symmetry can destroy the quantization of local Chern numbers, compromising the existence of edge modes. We observed non-integer local Chern number peaks that are non-zero inside the bulk bands but these non-zero non-integral local Chern numbers are not associated with the existence of robust edge states.

## 1. Introduction

The field of topological insulators has been extensively studied in the past decades. One of the most notable features of topological insulators is the topological edge/ boundary states that make the system conducting on the edge/boundary while insulating in the bulk [1-5]. It is believed that the topological edge/boundary states are robust against disorder under the condition that preserves the required symmetry. The effect of disorder on topological systems is itself a very interesting theoretical problem [6]. While disorder usually compromises useful phenomena, it has been proposed and demonstrated that the introduction of disorder can cause the transition from a topologically trivial phase to a nontrivial phase in the class of systems called topological Anderson insulator [7-10]. The effects of disorder on topological systems are subtle and complicated and should be considered on a case-by-case basis.

One challenge in this research field is how to evaluate the topological characteristics when systems are disordered or strongly perturbed. Due to the presence of disorder, the system is no longer periodic and hence the Chern number based on the Bloch band theory in momentum space is no longer applicable. There are many attempts in this regard, such as the covariant real-space formula for disordered AIII class in 1D [11], level spacing statistics of the entanglement spectrum [12], Bott index [13,14] and local Chern number [15-17]. Among them, the local Chern number is an index indicating the topological characteristics in the real space of a finite-sized sample under a chosen cutoff frequency.

In this work, we study the effects of disorder on topological transmission-line networks using local Chern numbers as a topological indicator. The networks are interesting signal transmission systems on their own and they are also very useful simulators for almost any kind of Hamiltonians in any dimension. For example, the Haldane model [18] can be simulated using network systems. In our system, the connectivity of the network ensures that the Hamiltonian can be block-diagonalized into three independent blocks and each block corresponds to an angular momentum eigenvalue. We will show that in a nonzero angular momentum subspace, the system can possess a nontrivial topological band gap, which behaves like the Haldane model. The chiral connectivity of the network induces non-zero Chern numbers in the bands within the subspaces with nonzero angular momentum. To introduce the disorder, we attach each node of the network with an isolated segment of cable so as to change the scattering at that node, which is equivalent to changing the onsite energy of a tight-binding model. Thus, the scattering at each node of the network becomes random when each segment has a random length. It is important to note that our specific disorder does not change the angular momentum of the original states. If the angular momentum is not conserved, then the states will be scattered to other angular momentum subspaces and the nontrivial topological property will be destroyed. By calculating the local Chern numbers and field patterns, we verify the robustness of the topological states under onsite disorder. The topological edge states are found to survive even when the disorder is strong enough to close the bulk band gap. We further study the case when the parity symmetry of the network is broken in the sense of the effective 2D space. We found non-integer local Chern number peaks inside the bulk bands. However, these peaks are found to be topologically trivial (not related to edge modes) and can be removed by adding disorder.

## 2. Tight-binding model

We will consider a network system that exhibits topological bands when periodic boundary conditions are applied to the network which has been used experimentally to realize angular momentum-dependent topological transport [17]. In order to explain why Chern numbers can arise in such a network without imposing an external magnetic field to break time-reversal symmetry, let us consider a tight-binding model as shown in Fig. 1(a). To be concise, Fig. 1(a) only shows one single hexagon of a periodic honeycomb lattice. There are six sublattices (black balls) $A_1$, $A_2$, $A_3$, $B_1$, $B_2$, $B_3$ in the hexagon, and they are distributed on three layers. Here, the subscripts denote the layer numbers. They constitute two meta-atoms $A$ and $B$, each carrying 3 nodes. The black and red lines denote the intra-layer and inter-layer couplings between the sub-lattices. The inter-layer couplings are connected in a chiral manner which has been used in realizing Weyl semimetals [19,20]. The couplings between layers need to be the same and the last layer is connected back to the first layer (head-to-tail connection) so that the model is periodic along the "vertical" direction. Due to the cyclic condition imposed along the "vertical" direction, a meta-atom can exhibit well-defined angular momentum eigenmodes and there are three angular momentum subspaces in which the phase differences between the neighboring layers are 0,

$\pm 2\pi/3$ respectively. Let us now "flatten" this 3D network (Fig. 1(a)) onto a 2D plane as shown in Fig. 1(b). After the "flattening", the atoms along the original "vertical" direction become sub-lattice sites inside a meta-atom on the 2D plane. For example, the atoms $A_1$, $A_2$, $A_3$ that belong to 3 different layers in Fig. 1(a) become the three sublattice nodes inside a meta-atom ("$A$") on a hexagonal 2D lattice. Likewise, $B_1$, $B_2$, $B_3$ in Fig. 1(a) become sublattice nodes of meta-atom $B$ in Fig. 1(b). It is this internal degree of freedom (or curled-up dimension) inside a meta-atom that allows for the possibility of a synthetic gauge flux defined through twisted coupling between the nodes. The network in Fig. 1(b) can be viewed as an effective 2D model in a fixed angular momentum subspace as we will see in the following.

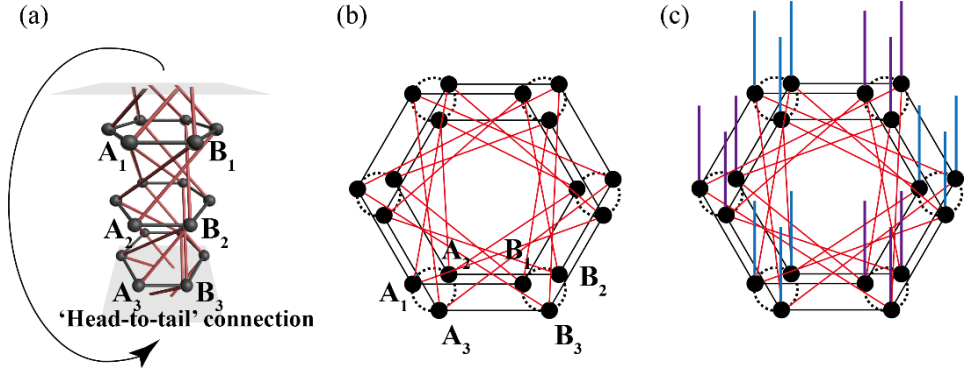

Figure 1. A ball-and-stick model that illustrates the connectivity of the network. A tight-binding model can be defined when hopping constants are specified for the bonds connecting the sites. (a) A 3D view of the tight-binding model. The $A$ sites and $B$ sites are sublattice sites of a hexagonal lattice in the $x$-$y$ directions. (b) The 3D model is compressed onto a 2D plane and is realized by transmission lines. The dashed circles mean that a meta-atom is composed of three nodes. (c) Additional transmission lines are added to the nodes to tune the onsite energies.

Following Ref. [17], the momentum space Hamiltonian $H_0$ of a tight-binding model of the periodic network system as depicted in Fig. 1(b) under the basis $(|A_1\rangle, |B_1\rangle, |A_2\rangle, |B_2\rangle, |A_3\rangle, |B_3\rangle)$ can be written as

$$H_0(\mathbf{k}) = \begin{pmatrix} 0 & t_2\beta_2 & t_1\beta_1 & 0 & t_1\beta_1^* & 0 \\ t_2\beta_2^* & 0 & 0 & t_1\beta_1^* & 0 & t_1\beta_1 \\ t_1\beta_1^* & 0 & 0 & t_2\beta_2 & t_1\beta_1 & 0 \\ 0 & t_1\beta_1 & t_2\beta_2^* & 0 & 0 & t_1\beta_1^* \\ t_1\beta_1 & 0 & t_1\beta_1^* & 0 & 0 & t_2\beta_2 \\ 0 & t_1\beta_1^* & 0 & t_1\beta_1 & t_2\beta_2^* & 0 \end{pmatrix}, \quad (1)$$

where

$$\beta_1 = 2\cos(\frac{3}{2}k_x a)\cdot\exp(i\frac{\sqrt{3}}{2}k_y a) + \exp(-i\sqrt{3}k_y a),$$

$$\beta_2 = 1 + 2\cos(\frac{\sqrt{3}}{2}k_y a)\cdot\exp(-i\frac{3}{2}k_x a). \quad (2)$$

Here $t_1$ and $t_2$ denote, respectively, the strengths of the interlayer and intralayer hoppings, $\mathbf{k}=(k_x,k_y)$ is the Bloch wavevector, and $a$ is the distance between the two meta-atoms. For now, the on-site energy is set to zero.

The parity symmetry can be broken by setting the on-site energy on meta-atoms $A$ and $B$ to be different with a difference of $2\Delta$. The symmetry-broken Hamiltonian can be written as

$$H(\mathbf{k}) = \begin{pmatrix} 0 & t_2\beta_2 & t_1\beta_1 & 0 & t_1\beta_1^* & 0 \\ t_2\beta_2^* & 0 & 0 & t_1\beta_1^* & 0 & t_1\beta_1 \\ t_1\beta_1^* & 0 & 0 & t_2\beta_2 & t_1\beta_1 & 0 \\ 0 & t_1\beta_1 & t_2\beta_2^* & 0 & 0 & t_1\beta_1^* \\ t_1\beta_1 & 0 & t_1\beta_1^* & 0 & 0 & t_2\beta_2 \\ 0 & t_1\beta_1^* & 0 & t_1\beta_1 & t_2\beta_2^* & 0 \end{pmatrix} + \Delta\cdot\begin{pmatrix} 1 & 0 & 0 \\ 0 & 1 & 0 \\ 0 & 0 & 1 \end{pmatrix}\oplus\begin{pmatrix} 1 & 0 \\ 0 & -1 \end{pmatrix},$$

$$(3)$$

Now we block-diagonalize $H(\mathbf{k})$ using the angular momentum basis through a unitary transform $\tilde{H}(\mathbf{k}) = U^{-1}H(\mathbf{k})U$, where

$$U = \frac{1}{\sqrt{3}}\begin{pmatrix} 1 & e^{\frac{2}{3}\pi i} & e^{-\frac{2}{3}\pi i} \\ 1 & 1 & 1 \\ 1 & e^{-\frac{2}{3}\pi i} & e^{\frac{2}{3}\pi i} \end{pmatrix}\otimes\begin{pmatrix} 1 & 0 \\ 0 & 1 \end{pmatrix}. \quad (4)$$

The block-diagonalized Hamiltonian becomes

$$\tilde{H}(\mathbf{k}) = \begin{pmatrix} -\mathrm{Re}(t_1\beta_1) + \sqrt{3}\,\mathrm{Im}(t_1\beta_1) & t_2\beta_2 \\ t_2\beta_2^* & -\mathrm{Re}(t_1\beta_1) - \sqrt{3}\,\mathrm{Im}(t_1\beta_1) \end{pmatrix} \oplus \begin{pmatrix} 2\mathrm{Re}(t_1\beta_1) & t_2\beta_2 \\ t_2\beta_2^* & 2\mathrm{Re}(t_1\beta_1) \end{pmatrix}$$

$$\oplus \begin{pmatrix} -\mathrm{Re}(t_1\beta_1) - \sqrt{3}\,\mathrm{Im}(t_1\beta_1) & t_2\beta_2 \\ t_2\beta_2^* & -\mathrm{Re}(t_1\beta_1) + \sqrt{3}\,\mathrm{Im}(t_1\beta_1) \end{pmatrix} + \Delta \cdot \begin{pmatrix} 1 & 0 & 0 \\ 0 & 1 & 0 \\ 0 & 0 & 1 \end{pmatrix} \otimes \begin{pmatrix} 1 & 0 \\ 0 & -1 \end{pmatrix}$$

. (5)

Equation (5) can also be expressed as

$$H(\mathbf{k}) = f_x(\mathbf{k})\sigma_x + f_y(\mathbf{k})\sigma_y + f_0(\mathbf{k})\cos\left(\frac{2m\pi}{3}\right)\sigma_0 + f_z(\mathbf{k})\sin\left(\frac{2m\pi}{3}\right)\sigma_z$$

$$+ \Delta \cdot \sigma_z \; , (6)$$

where $\sigma_0$ is the $2\times 2$ identity matrix, $\sigma_x$, $\sigma_y$, and $\sigma_z$ are the Pauli matrices. $m$ denotes three subspaces, each with a fixed angular moment of 0, +1 or -1. In the following study, we choose the angular momentum $m=1$ subspace for the consideration of disorder. The functions $f_0(\mathbf{k})$, $f_x(\mathbf{k})$, $f_y(\mathbf{k})$ and $f_z(\mathbf{k})$ have the following forms:

$$f_0(\mathbf{k}) = t_1[4\cos(\tfrac{3}{2}k_x a)\cos(\tfrac{\sqrt{3}}{2}k_y a) + 2\cos(\sqrt{3}k_y a)],$$

$$f_x(\mathbf{k}) = t_2[1 + 2\cos(\tfrac{\sqrt{3}}{2}k_y a)\cos(\tfrac{3}{2}k_x a)],$$

$$f_y(\mathbf{k}) = 2t_2\cos(\tfrac{\sqrt{3}}{2}k_y a)\sin(\tfrac{3}{2}k_x a),$$

$$f_z(\mathbf{k}) = t_1[4\cos(\tfrac{3}{2}k_x a)\sin(\tfrac{\sqrt{3}}{2}k_y a) - 2\sin(\sqrt{3}k_y a)].$$

(7)

The Eq. (6) is a $2\times 2$ Hamiltonian in the angular momentum subspace $m$. The first two terms of Eq. (6) give Dirac cone dispersions at the K and K' points of the Brillouin zone, which are typically those of a hexagonal lattice. The third term induces a global energy shift, which does not change the band topology. The fourth term depends on the angular momentum index "$m$" and will gap the system and give rise to Chern numbers in the bands. The Chern numbers arise from the synthetic gauge field felt by the waves traveling with a well-defined angular momentum index "*m*". It can be viewed as an analog of the spin-orbital coupling term in the quantum spin Hall effect, except that the spin is now replaced by the orbital angular momentum of the waves. The parity-breaking parameter Δ will also gaps the system, but induces a topologically trivial gap.

## 3. Transmission line network

A transmission line is a one-dimensional waveguide and by connecting transmission lines into a network, the network system can mimic different kinds of tight-binding models. In addition, prior results show that simulated results using network equations (see below) are almost identical to measured results in experiments [17] and hence a numerical simulation can predict with confidence the properties of the real system. This is in contrast to photonic crystal systems in which the multiple scattering among meta-atoms makes the finite range hopping models an approximation at best. In the network system, the voltages on the nodes are determined by the network equations:

$$-\psi_i \sum_j \coth(gl_{ij}) + \sum_j \frac{1}{\sinh(gl_{ij})} \psi_j = 0. \tag{8}$$

Here, $\psi_i$ is the voltage at the *i*-th node, $l_{ij}$ is the length of the cable connecting nodes i and j, and $g = (i\omega/c_0)\sqrt{\varepsilon}$ with $\omega$, $c_0$ and $\varepsilon$ being the angular frequency, the speed of light, and the relative permittivity of the dielectric medium in the transmission lines, respectively. The network equation can be mapped to a tight-binding model (with energy equal to zero), where $-\sum_j \coth(gl_{ij})$ is an on-site term and the $1/\sinh(gl_{ij})$ coefficient can be mapped to a hopping term. The nodes can be conceptually mapped to "meta-atoms" or internal degrees of freedom of a meta-atom and transmission lines provide the "hopping" between "atoms". We note that the coupling coefficient in the network depends on the frequency, and hence the network is more complex and richer in physics than a standard tight-binding model whose coupling coefficients are constants. Unless otherwise stated, the lengths of intralayer and interlayer cables are taken to be 0.43 m and 2.06 m in the simulations.

In the following, we will skip the length unit 'meter' for simplicity. We set the wave speed inside the cable to $0.66c_0$, according to experimentally determined values [17]. For simplicity, the loss in the cable is ignored in the simulation but can be included if we wish. The network shown in Fig. 1(b), with network parameters as specified above, has a nontrivial band gap between 31 MHz and 36.3 MHz for the $m = \pm 1$ subspaces.

To change the on-site terms of the network equation without changing the hopping terms, we can directly connect an additional cable to each node, which changes the scattering characteristics of the node. For example, if we connect a cable to the node $i$, the network equation for that node becomes:

$$-\psi_i \sum_j \coth(gl_{ij}) + \sum_j \frac{1}{\sinh(gl_{ij})} \psi_j - \psi_i \coth(gl_{iC}) + \frac{1}{\sinh(gl_{iC})} \psi_C = 0, \quad (9)$$

where node $C$ denotes the other end point of the additional cable and $l_{iC}$ is the length of this cable. The summation is over all the neighboring nodes except the additional node $C$. The last two terms are additional terms that appear after adding the cable. For the end node $C$, its network equation is:

$$-\psi_C \coth(gl_{iC}) + \frac{1}{\sinh(gl_{iC})} \psi_i = 0. \quad (10)$$

We can eliminate $\psi_C$ by substituting Eq. (10) into Eq. (9) and obtain:

$$\psi_i \left[ -\sum_j \coth(gl_{ij}) - \tanh(gl_{iC}) \right] + \sum_j \frac{1}{\sinh(gl_{ij})} \psi_j = 0. \quad (11)$$

Equation (11) shows that from the viewpoint of a tight-binding model, the onsite term can be controlled by changing the length of the additional cable without altering the hopping terms. We emphasize here that all the simulation results that will be shown later are obtained by solving these network equations explicitly. The tight-binding equations in Section 2 are just for understanding why Chern numbers can exist in such a system defined by connectivity.

## 4. Disordered networks

In this section, we will introduce onsite disorder to the network and study its influence on the topological properties in the subspace $m=1$, which exhibits bands with non-zero Chern numbers and is hence topologically non-trivial. The onsite disorder is achieved by adding additional cables on the nodes as explained in the previous section. To preserve the angular momentum $m=1$, we need to add three additional cables to the three nodes of the same meta-atom as shown in Fig. 1(c). The lengths of these three cables of one meta-atom are set to be the same (Fig. 1(c)) so that the network is still invariant within the curled-up dimension in the 2D model. While the additional cables are the same on the nodes inside one meta-atom, they can be different at different meta-atoms if we want to break parity as indicated by the blue and purple lines in Fig. 1(c) representing the additional cables on sublattices $A$ and $B$, respectively. We label their lengths as $L_A$ and $L_B$.

We start with an ordered system without any onsite cables. We then introduce randomness to the system by attaching the onsite cables of random lengths at each meta-atom and the length of cables on each meta-atom is uniformly distributed in the range $[0, L_r]$. We first consider a finite network containing 9 rows and 9 columns, and we calculate the frequency spectrum as a function of the disorder strength parameter $L_r$ for the subspace of $m=1$ by showing the network equations. The results are shown in Fig. 2(a). We found that a non-trivial band gap, which is marked by a red curve, survives up to a certain value of the disorder parameter. By examining the wavefunctions, we find that the states inside the gap (see fig. 2(c)) are topological edge states that decay exponentially into the bulk (note the intensity map in Fig. 2(c) is color-coded in log-scale). This gap closes finally with the increase of $L_r$. The purple dashed line corresponds to the case of a small disorder of $L_r=0.1$.

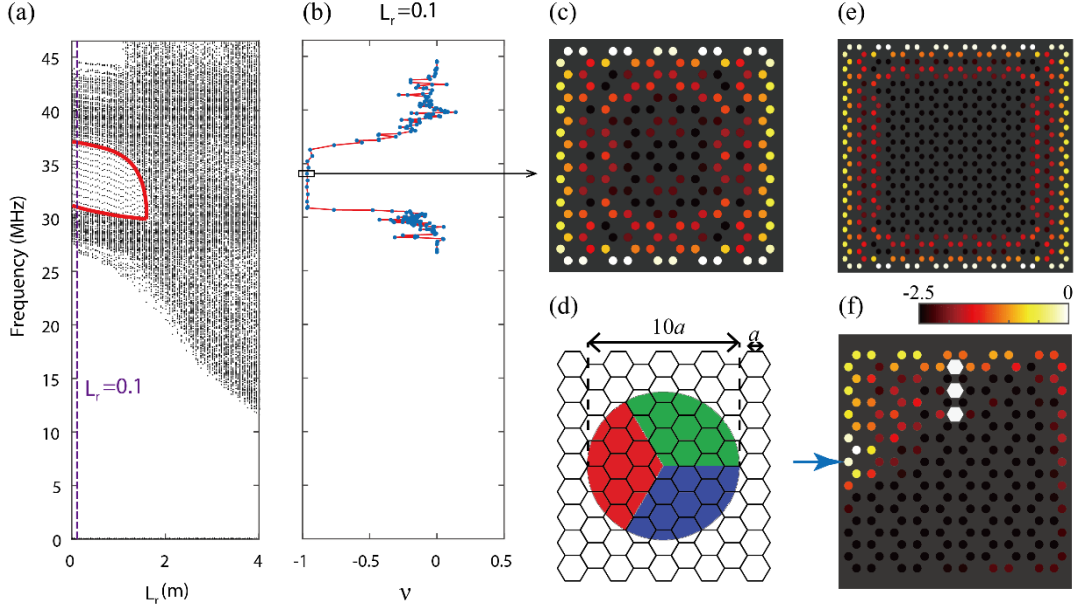

Figure 2. (a) Energy spectrum vs. the disorder strength parameter $L_r$. The red line marks the topological non-trivial gap which closes gradually as $L_r$ increases. The states inside the gap are localized on the edges. (b) The computed local Chern numbers as a function of frequency for $L_r=0.1$, which is marked by the dotted line in panel (a). In the gap region, the local Chern number is close to -1. (c) The field pattern of an edge state in the gap marked by the arrow in (b). (d) The computational domain is a 9×9 finite-sized sample. The red/blue/green highlights the 3 sectors used in the local Chern number computation. (e) The edge state pattern of a larger sample with 15 rows by 15 columns for $L_r=0.1$. (f) The one-way propagating edge wave in a sample with an inserted defect.

To characterize the topological property of this random network, we plot the local Chern number as a function of the cutoff frequency ($f_c$) for the disorder strength of $L_r=0.1$ as shown in Fig. 2(b). The local Chern number is defined by the anti-symmetric product of the projection operators:

$$v(r) = 12\pi i \sum_{j \in U} \sum_{k \in V} \sum_{l \in W} (P_{jk} P_{kl} P_{lj} - P_{jl} P_{lk} P_{kj}), \quad (12)$$

where $P = \sum_{f \leq f_c} |u_f\rangle\langle u_f|$ is the projection operator which adds up all the eigenstates $|u_f\rangle$ below a cutoff frequency $f_c$, and $P_{ij} = \langle x_i | P | x_j \rangle$ gives the spatial correlation

between sites $x_i$ and $x_j$ in different sectors. These sites lie in three different sectors ($U$, $V$, and $W$ indicated by red, blue and green) of the computational domain as shown in Fig. 2(d). Then $v(r)$ is taken as the local Chern number of the computational domain center. Our local Chern number curves refer to the local Chern number computed at the sample center, and the radius of the computational domain is chosen to be $r=5a$.

In Fig. 2(b), we see that the local Chern number of the disordered system is close to 0 in the bulk band continuum energy regimes and approximately -1 in the band gap region, indicating that the band gap remains topologically non-trivial in the presence of the disorder. We plot the voltage strength field pattern of a topological edge state inside the gap as shown in Fig. 2(c). In each angular momentum subspace, the field amplitudes in each of the three nodes inside a meta-atom differ only by a phase factor. Therefore, we only need to plot the field pattern of one of the nodes. We also check other states inside the gap and find that all of them are topological edge states. To show that the topological property does not depend on the sample size, we calculate an edge state of a larger sample (15×15) for the case of $L_r = 0.1$ as shown in Fig. 2(e). We see that the edge states are more localized to the edge, as expected. Furthermore, we introduce the defect on the edge to check the one-way propagating property of the edge states. The defect is introduced by removing all the inter-layer cables in the three unit-cells marked by white hexagons as shown in Fig. 2(f). The signal with 32.7MHz is injected from the meta-atom pointed by the blue arrow. In order to see the propagation direction, the loss is considered in this simulation and the absorption length is assumed to be $L \cong 338 \times f^{-0.6123}$ and $g \approx ik - 1/2L$. In Fig. 2(f), it can be clearly seen that the edge wave propagates around the defect without backscattering.

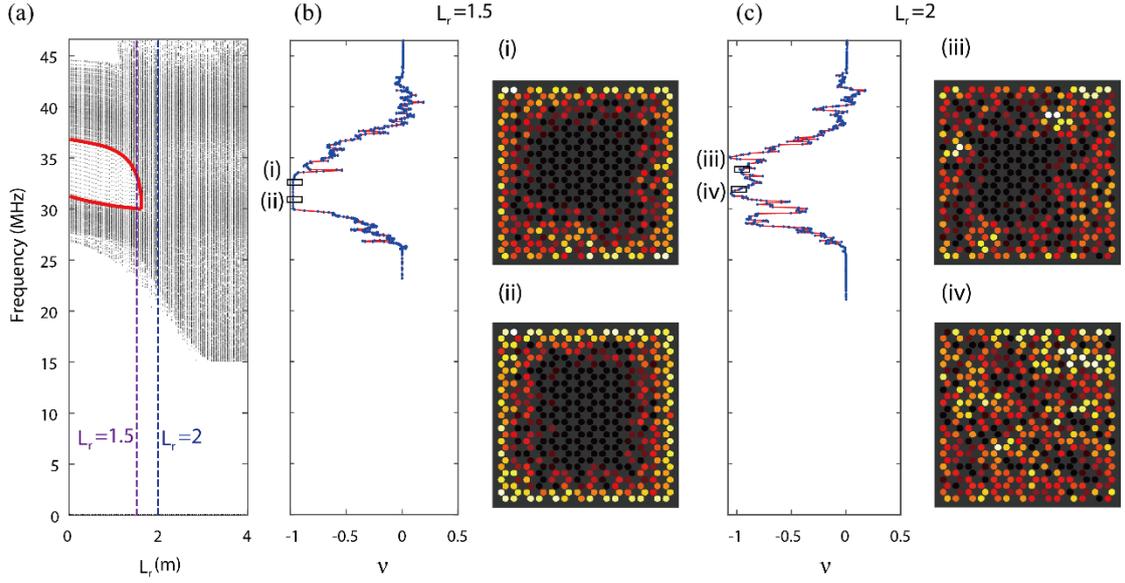

Figure 3. (a) Energy spectrum vs. $L_r$ for a 13×13 sample. (b) The left panel is the computed local Chern numbers as a function of frequency for $L_r=1.5$. The right panel shows two edge states with the frequencies marked in the left panel. (c) The left panel is the local Chern numbers for a bigger disorder of $L_r=2$. The right panel shows a bulk state and an edge state, respectively.

In the following, we increase the randomness by increasing the value of $L_r$ and plot the local Chern numbers for $L_r=1.5$ and $L_r=2$ as shown in Fig. 3(b) and 3(c). Here we use a sample of 13×13. For $L_r=1.5$, the nontrivial gap still exists and the local Chern number approaches -1 in the gap regime. Two field patterns of the states in the gap are shown in Fig. 3(b) (i) and (ii). These states decay exponentially into the bulk, and so they are edge modes. When $L_r=2$, we see that the gap already closes (Fig. 3(a)). However, the local Chern number still approaches -1 in a frequency region between 30MHz and 35MHz as shown in 3(c), although there are conspicuous fluctuations. We check the field patterns in this regime and find that some of the modes still retain edge mode characters and they are localized in near boundary, while other modes have penetrated into the bulk. Figure 3(c) (iii) and (iv) show the two examples (basically chosen at random). This tells us that the closure of the gap indicates the merging of the bulk states with the edge states in the frequency domain. The closure of the gap does not immediately indicate the disappearance of the local Chern number and edge states. It indicates the robustness of the topological edge states against the onsite disorder.

## 5. Breaking parity symmetry

In this section, we will break the parity symmetry of the network by using different additional cable lengths on sublattices $A$ and $B$ to see its effects on the topological properties. In other words, the lengths of purple and blue cables in Fig. 1(c) are different. We note that this parity symmetry is in the sense of the effective 2D model as depicted in Fig. 1(b). We first fix the length $L_B$=1.9 and calculate the energy spectra as a function of $L_A$. The results are shown in Fig. 4(a). Here, the sample size is 9×9.

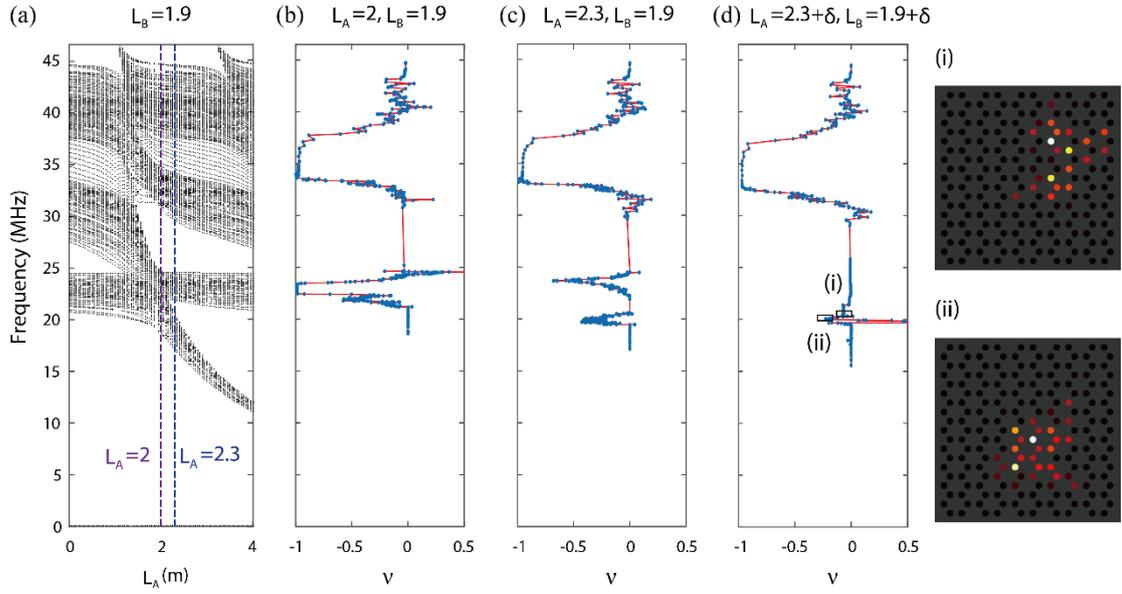

Figure 4. (a) Energy spectrum as a function of $L_A$. The $L_B$ is fixed at $L_B$=1.9. (b)-(d) The local Chern number curves for the three cases: (b) $L_A$=2, $L_B$=1.9, (c) $L_A$=2.3, $L_B$=1.9, (d) $L_A = 2.3 + \delta$, $L_B = 1.9 + \delta$.

We first consider the case of $L_A$=2. In this case, the parity is only slightly broken as $L_B$=1.9. The calculated local Chern number as a function of cutoff frequency is shown in Fig. 4(b). From Fig. 4(b) we see there are two nontrivial bulk band gaps where the local Chern number approaches -1. By examining the wavefunction, we found by all the states inside the two nontrivial gaps are topological edge states. Here, we focus on the lower frequency gap, which is smaller and more sensitive to system parameter changes. We increase the parity symmetry breaking strength by setting $L_A$=2.3 and $L_B$=1.9. The corresponding local Chern numbers are shown in Fig. 4(c). We see that the lower nontrivial gap now becomes a trivial gap and no edge states exist in this gap. However, it is found that the local Chern number curve has two peaks in the frequency ranges of the two bulk bands, but no topological edge states are found at these frequencies. This unusual phenomenon of non-integral but non-zero Chern number can be explained as follows. The local Chern number only includes the contributions of the states under the cutoff frequency. Therefore, the local Chern number peaks correspond to the integrated Berry curvature of all the states under the cutoff frequencies, but not the entire bulk band. Due to the parity symmetry breaking, the Berry curvatures are not zero everywhere in the Brillouin zone. However, the Berry curvature integration of a whole band is still zero as the gap is trivial. It can be verified that the local Chern number is zero at the frequency of the bulk band edge. The local Chern number peaks originated from the bulk bands are not robust that they can be destroyed by a change of parameters. To verify this, we introduce the onsite disorder by setting $L_A = 2.3 + \delta$ and $L_B = 1.9 + \delta$, where $\delta$ is a random number uniformly distributed in [0,0.5]. The corresponding local Chern number curve is shown in the left panel of Fig. 4(d). Due to the disorder, the peaks disappeared. We examine the states in this frequency regime and find that they are all localized bulk states. The right panel of Fig. 4(d) shows two examples of such localized states.

## 6. Conclusion

In short, we studied the topological properties of transmission line networks under the influence of onsite disorder and parity symmetry breaking. When the onsite disorder is introduced, the topological band gap will gradually close. However, the closure of the gap does not immediately destroy all the topological edge states. From the local Chern number curves and field patterns, we found that some topological edge states still survive after the closure of the band gap. When the parity symmetry is broken, the nontrivial band gap will transit to a trivial gap when the symmetry breaking is strong enough. After the transition, local Chern number peaks emerge in the bulk bands. However, these peaks do not correspond to integer values and are not associated with topological edge states. They are also sensitive to details and are easily destroyed by disorder.

**Acknowledgment:** We thank Prof. Z.Q. Zhang for many helpful discussions. This work is supported by Research Grants Council (RGC) Hong Kong through grant 16307420.

# References

1. C. L. Kane, E. J. Mele, Physics. A new spin on the insulating state, Science (New York, NY) **314,**1692 (2006)
2. X.-L. Qi, S.-C. Zhang, The quantum spin Hall effect and topological insulators, Physics Today **63,**1 (2010)
3. M. Z. Hasan, C. L. Kane, Colloquium: topological insulators, Reviews of Modern Physics **82,**3045 (2010)
4. X.-L. Qi, S.-C. Zhang, Topological insulators and superconductors, Reviews of Modern Physics **83,**1057 (2011)
5. A. B. Khanikaev, S. H. Mousavi, W. K. Tse, M. Kargarian, A. H. MacDonald, G. Shvets, Photonic topological insulators, Nature Materials **12,**223 (2013)
6. E. Prodan, Disordered topological insulators: a non-commutative geometry perspective, Journal of Physics A: Mathematical and Theoretical **44,**113001 (2011)
7. J. Li, R. L. Chu, J. K. Jain, S. Q. Shen, Topological Anderson insulator, Physical Review Letters **102,**136806 (2009)
8. H. M. Guo, G. Rosenberg, G. Refael, M. Franz, Topological Anderson insulator in three dimensions, Physical Review Letters **105,**216601 (2010)

9. C. W. Groth, M. Wimmer, A. R. Akhmerov, J. Tworzydło, C. W. J. Beenakker, Theory of the topological Anderson insulator, Physical Review Letters **103,**196805 (2009)
10. S. Stützer, et al., Photonic topological Anderson insulators, Nature **560,**461 (2018)
11. I. Mondragon-Shem, T. L. Hughes, J. Song, E. Prodan, Topological criticality in the chiral-symmetric AIII class at strong disorder, Physical Review Letters, **113,**046802 (2014)
12. E. Prodan, T. L. Hughes, B. A. Bernevig, Entanglement spectrum of a disordered topological chern insulator, Physical review letters **105,**115501 (2010)
13. H. Huang, F. Liu, Quantum spin Hall effect and spin Bott index in a quasicrystal lattice, Physical Review Letters **121,**126401 (2018)
14. X. S. Wang, A. Brataas, R. E. Troncoso, Bosonic Bott index and disorder-induced topological transitions of magnons, Physical Review Letters **125,** 217202 (2020)
15. N. P. Mitchell, et al., Amorphous topological insulators constructed from random point sets, Nature Physics **14,**380 (2018)
16. A. Kitaev, Anyons in an exactly solved model and beyond, Annals of Physics **321,**2 (2006)
17. T. Jiang, M. Xiao, W. J. Chen, L. Yang, Y. Fang, W. Y. Tam, C. T. Chan, Experimental demonstration of angular momentum-dependent topological transport using a transmission line network, Nature communications **10,**1 (2019)
18. F. D. M. Haldane, Model for a quantum Hall effect without Landau levels: Condensed-matter realization of the "parity anomaly", Physical Review Letters **61,**2015 (1988)
19. M. Xiao, W.-J. Chen, W. Y. He, C. T. Chan, Synthetic gauge flux and Weyl points in acoustic systems. Nature Physics, **11,**920 (2015)
20. W.-J. Chen, M. Xiao, C. T. Chan, Photonic crystals possessing multiple Weyl points and the experimental observation of robust surface states, Nature Communications **7,**1 (2016)